\documentstyle[psfig]{mn}
	\title[R-mode instability of rapidly rotating neutron stars]{A numerical study of the r-mode instability of rapidly rotating nascent neutron stars}
	\author[Yoshida et al.]{Shin'ichirou Yoshida${}^1$\thanks{present address: Scuola Internazionale Superiore di Studi Avanzati, Via Beirut 2-4, 34014 Trieste, Italia}
, Shigeyuki Karino${}^1$, Shijun Yoshida${}^2$ and Yoshiharu Eriguchi${}^1$\\
${}^1$Department of Earth Science and Astronomy,
Graduate School of Arts and Sciences, 
University of Tokyo, \\
Komaba, Meguro-ku, Tokyo 153-8902, Japan\\
${}^2$Astronomical Institute, 
Graduate School of Science, 
Tohoku University, 
Sendai 980-8578, 
Japan }
	\date{Accepted, Received}
\begin{document}
	\maketitle
\begin{abstract} 
First results of numerical analysis of classical r-modes 
of {\it rapidly} rotating 
compressible stellar models are reported. The full set of linear perturbation
equations of rotating stars in Newtonian gravity are numerically solved 
without the slow rotation approximation. A critical curve of gravitational
wave emission induced instability which restricts the rotational frequencies 
of hot young neutron stars is obtained. Taking the standard cooling mechanisms 
of neutron stars into account, we also show the `evolutionary curves' along 
which neutron stars are supposed to evolve as cooling and spinning-down 
proceed.  Rotational frequencies of $1.4M_{\odot}$ stars suffering from this 
instability decrease to around $100$Hz when the standard cooling mechanism of 
neutron stars is employed. This result confirms the results of other authors 
who adopted the slow rotation approximation.

\end{abstract}
	\begin{keywords}
	stars: neutron -- stars: rotation -- stars: oscillations
	\end{keywords}
	
%%%%%--------------------
\section{Introduction}
Recently the r-mode instability induced by gravitational wave emission 
was discovered \cite{andersson,friedman-morsink}. It is an axial mode 
variant of the secular instability of stellar oscillations excited by 
coupling with a radiation field \cite{chandrasekhar,friedman-schutz}.

The r-mode oscillations are the extension of Rossby-Haurwitz 
wave of a rotating fluid known in geophysical literatures to global stellar 
oscillations \cite{papaloizou-pringle}. As they are dominated by the 
tangential velocity perturbations, their coupling with gravitational 
radiation was once considered to be weaker than those of the polar modes 
(e.g. f-modes). Thus it was so striking and stimulating to the astrophysical 
community that many people began to study whether or not the instability may 
be strong enough to set a severe limit on the rotational period of neutron 
stars.  In particular, several authors 
\cite{lindblom-owen-morsink,owen,andersson-kokkotas-schutz,lockitch-friedman,lindblom-mendell-owen,yoshida-lee}
have investigated its effect on the evolution of the stellar rotational 
frequencies (see Friedman \& Lockitch~\shortcite{friedman-lockitch} for a 
recent review). 
Those works indeed have revealed that the instability is strong enough
to limit severely the rotational frequencies of hot young neutron stars 
which are born with the initial temperature of $\sim 10^{11}$K. Even if the
neutron stars are born with the nearly Keplerian (mass-shedding limit) rotation
frequency, they settle down to states with only a small fraction of their 
original angular velocity, i.e. with less than $10~\%$ of the Keplerian 
rotational frequency, when they cool down to the neutron superfluid transition 
temperature, $T_c \sim 10^9$K. 
As for the relatively cold and old neutron stars as those in X-ray binaries 
or radio pulsars, this instability may also set a severe limit on their 
rotational frequencies \cite{bildsten,andersson-kokkotas-stergioulas,levin}.

Thus far, these results have been obtained by extrapolating the results of the 
slow rotation approximation \cite{papaloizou-pringle,provost,saio} to rotating 
models with the Keplerian frequency, except 
that Lindblom and Ipser~(1999) obtained analytically
the eigenfrequencies and the eigenfunctions of the 'classical' as well 
as the 'generalized' r-modes for the Maclaurin spheroids.
Moreover, Bildsten \& Ushomirsky~\shortcite{bildsten-ushomirsky}
recently studied a damping effect of a viscous boundary layer formed 
at the stellar crust--core interface. According to their work, the viscous 
boundary layer stabilizes the system significantly and the upper limit of 
the rotational frequency of neutron stars may become around $\sim 500$Hz. 
If this is the case, the mode analysis without the slow rotation approximation 
should be needed.

Thus in order to see whether the 
evolutionary picture mentioned above is quantitatively correct and to 
investigate the evolution of stellar spins, it is indispensable to study the
r-modes by taking rapid rotation into account.

In this {\it Letter}, we will show the first numerical results of the 
`classical' r-modes for rapidly rotating stars. Here classical r-modes are 
oscillations with $l=m$, where $l$ and $m$ are the indices of spherical 
harmonics. 
Our numerical scheme is the improved version of the one used in the polar mode 
analysis \cite{yoshida-eriguchi} and will be published elsewhere together with 
more detailed results for rapidly rotating compressible stars
\cite{karino-yoshida-yoshida-eriguchi}.

%%%%%---------------
\section{Evaluation of damping times}

Once eigenfrequencies and eigenfunctions are obtained, timescales of the
change of the system due to gravitational radiation and viscosity can be 
estimated by the standard procedure \cite{lindblom-owen-morsink}.

The canonical energy $E_c$ of the perturbation is defined as:
\begin{equation}
E_c \equiv  \frac{1}{2}\int d^3x \cdot\left[ \rho\delta \vec{v}\cdot
\delta\vec{v}^{~*}+ (\delta p/\rho-\delta\phi)^* \delta\rho\right],
\end{equation}
%
%\begin{eqnarray}
%E_c &=& \frac{1}{2}\int d^3x \cdot\left[ \rho\delta \vec{v}\cdot\delta\vec{v}^{~*}\right.
%\nonumber\\
%&+& \left.\frac{(\delta p/\rho-\delta\phi)^* \delta\rho
%+ (\delta p/\rho-\delta\phi) \delta\rho^*}{2}\right],
%\end{eqnarray}
%
where $\delta\rho, \delta p, \delta\vec{v}, \delta\phi$ are the 
Eulerian perturbations of the density, pressure, velocity and 
gravitational potential, respectively.
The superscript, $*$, means the complex conjugation of the corresponding
quantity. 

When we assume that the perturbed quantities behave as 
$\sim e^{-i(\omega t - m \varphi)}$, 
the energy dissipation rate due to gravitational radiation is expressed by 
the multipole radiation formula as follows:
\begin{equation}
\dot{E}_{gr} = (\omega-m\Omega)\sum_{l\ge m}N_l \omega^{2l+1}
     \left(\left|\delta D_{lm}\right|^2 + \left|\delta J_{lm}\right|^2\right),
\end{equation}
where $\delta D_{lm}$ and $\delta J_{lm}$ are the mass and the mass current
multipoles, respectively, defined as:
\begin{equation}
      \delta D_{lm} = \int d^3x \cdot\delta\rho r^l Y^*_{lm} ,
\end{equation}
\begin{equation}
      \delta J_{lm} = \frac{2}{c}\sqrt{\frac{l}{l+1}}
\int d^3x \cdot r^l (\rho\delta v + \delta\rho v)
\cdot\vec{Y}^{B*}_{lm} ,
\end{equation}
where $c$ is the speed of light.
Here the magnetic type vector spherical harmonics $\vec{Y}^{B}_{lm}$
is defined as:
\begin{equation}
\vec{Y}_{lm}^B = [l(l+1)]^{-\frac{1}{2}}\vec{r}\times \nabla Y_{lm} .
\end{equation}
The coupling constant $N_l$ is a function of $l$ as follows:
\begin{equation}
N_l = \frac{4\pi G}{c^{2l+1}}\frac{(l+1)(l+2)}{l(l-1)[(2l+1)!!]^2}.
\end{equation}

%% Apr.23
As the rapidly rotating stars are highly deformed from spherical 
configuration, the summation (2) in the index $l$ should include
theoretically the infinite numbers of multipole terms. We take 
the lowest five ones into account here for simplicity. 
It should be noted that the slow rotation approximation includes 
only the lowest order term.
On the other hand, the energy dissipation rate due to the fluid viscosity is
written as:
\begin{equation}
\dot{E}_s = -\int d^3x \cdot2\eta\delta\sigma^{ab} \delta\sigma_{ab}^* ,
\end{equation}
for the shear viscosity with a coefficient $\eta$, and
\begin{equation}
\dot{E}_b = -\int d^3x \cdot\zeta \left|\delta\Theta\right|^2 ,
\end{equation}
for the bulk viscosity with a coefficient $\zeta$.  Here the volume 
expansion rate $\delta\Theta$ and the shear tensor $\delta\sigma_{ab}$ are 
defined as:
\begin{equation}
    \delta\Theta \equiv \nabla_c\delta v^c,
\end{equation}
and
\begin{equation}
   \delta\sigma_{ab} \equiv \nabla_{\left(a\right.}\delta v_{\left.b\right)} -
   \frac{\delta\Theta}{3}\delta_{ab},
\end{equation}
where a round bracket means symmetrization among indices. These energy
dissipations and their rates are numerically evaluated for the 
rotating stellar models.

Then the imaginary part of the eigenfrequency of the r-mode, 
$\tau_r^{-1}$, is written as follows: 
\begin{equation}
\tau_r^{-1} = \frac{\dot{E}_{gr}+\dot{E}_s+\dot{E}_b}{2E_c}
\equiv \tau_{gr}^{-1}+\tau_s^{-1}+\tau_b^{-1}.
\end{equation}
Positive values of $\tau_r^{-1}$ correspond to the states where the r-mode 
instability dominates the stabilizing viscous effect. Consequently critical 
states of the instability are defined by $\tau_r^{-1}=0$.

%%%%%---------------
\section{Comparison with the slow-rotation approximation}
We have studied the adiabatic perturbations of the equilibrium sequence of 
`canonical neutron stars', i.e. polytropes with $p=K\rho^2$ whose mass and 
radius in the spherical limit become $1.4M_\odot$ and $12.5$km, respectively. 
Here $p$, $\rho$ and $K$ are the pressure, the density and a polytropic
constant, respectively.
This sequence is the same as that analyzed by many authors in the slow 
rotation approximation. In Fig~1, the eigenfrequency 
$\omega$ normalized
by the angular velocity $\Omega$ is plotted against the normalized angular
velocity $\Omega/\sqrt{4 \pi G \rho_c}$ where $\rho_c$ and $G$ are the 
central density and the gravitational constant, respectively. 
\footnote{We do not display the results of our computation in
the vicinity of the mass-shedding limit, because the eigenfunctions 
are rather ill-behaved and the estimate of the instability time-scale 
is difficult there. Our results (Fig.~1 and Fig.~2) 
are terminated at the model with 
$98.7$\% of the mass-shedding limit frequency ($850$Hz).
}
To compare our results with those by the slow rotation approximation, we also
plot the solutions by Yoshida and Lee \cite{yoshida-lee} 
where the third order term
in the $\Omega$ expansion is taken into account. The result of the 
slowest rotation in our results ($\Omega/\sqrt{4 \pi G \rho_c}=0.0154$)
agrees with theirs within the relative error of $0.4$\%.
The grid point numbers used in this computation are $32$ for the radial
coordinate $r$ from the center to the surface of the star, and $11$ for 
the angular coordinate $\theta$ in the quarter of the meridional 
cross section of the star.
%%%%%%%%%%%%%%%%%%%%%%%%%%%%%%%%%%%
%%%%FIG.1
\begin{figure}
        \centering\leavevmode
        \psfig{file=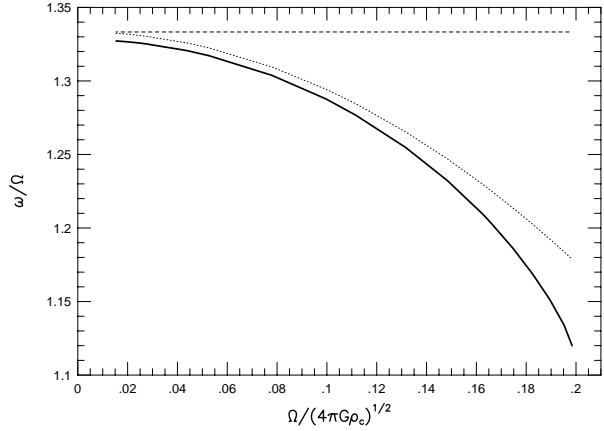,width=9.cm,angle=90,clip=}
        \caption{The normalized eigenfrequency $\omega/\Omega$ is
        plotted against the normalized angular velocity 
        $\Omega/(4 \pi G \rho_c)^{1/2}$. Also plotted are the solutions
	by the slow rotation approximation. The dashed line is the 
	solution of the first order in $\Omega$, whereas the dotted line
	is the one in which the third order correction is also taken into
	account. 
	The mass-shedding limit of this sequence is the point where 
	$\Omega/(4 \pi G \rho_c)^{1/2}=0.205$}
	\label{fig1}
\end{figure}
%%%%%%%%%%%%%%%%%%%%%%%%%%%%%%%%%%%

\section{Critical curve and 'evolutionary track' of a hot and young neutron star}

%%%%FIG.2
\begin{figure}
        \centering\leavevmode
        \psfig{file=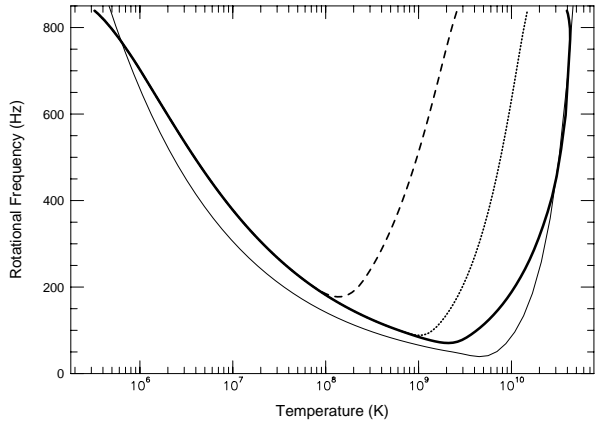,width=9.cm,angle=90,clip=}
        \caption{Critical rotational frequency of a $1.4M_\odot$ neutron 
	star with the polytropic equation of state $p = K \rho^2$ is 
	plotted against the temperature by the thick solid curve for the 
	$l=m=2$ mode. For comparison the critical curve computed using the 
	slow rotation approximation is drawn as well (thin solid line).
	Also shown are the curves on which the cooling 
	time-scale of the star is equal to the r-mode instability 
	time-scale (dotted line for the modified URCA process; dashed 
	line for the `quasi-particle' $\beta$-decay in pion condensation).}
	\label{fig2}
\end{figure}
%%%%%

As for the source of shear and bulk viscosity, we adopt the 
same microphysical process as that used by Ipser and Lindblom 
\cite{ipser-lindblom}; that is, if the stellar temperature T satisfies the
relation $T>T_c$, the dominant processes are neutron-neutron (nn) collision 
and electron-electron (ee) collision. They result in the shear viscosity 
coefficient with the following density and temperature dependence: 
\begin{equation}
\eta_{nn} \propto \rho^{9/4} T^{-2} ,
\end{equation}
and 
\begin{equation}
\eta_{ee} \propto \rho^{2} T^{-2} .
\end{equation}
When $T\le T_c$, only the electron collision 
is included. The bulk viscosity arises from lag of the 
$\beta$-reaction to the oscillation.  The viscosity coefficient has the 
following dependence:
\begin{equation} 
\zeta \propto \rho^2(\omega-m\Omega)^{-2}T^6 .
\end{equation}

For each mode, the inverse of the time-scale $\tau_r^{-1}$ depends on the 
stellar temperature $T$ as well as on the stellar rotational frequency $f$. 
In Fig~\ref{fig2}, we plot the critical points of stability on the $T-f$ plane
for the $l=m=2$ mode. Above this critical curve, gravitational wave emission 
instability dominates the stabilization effect due to viscosity. Along the 
fixed temperature line, i.e. vertical line, the instability time-scale 
decreases approximately as $\sim f^{2m+2}$.  As seen from this figure,
the smallest value of the rotational frequency of the critical curve
is roughly $8$ \% of the frequency at the mass-shedding limit. 

%% Apr.23
Also shown is the critical curve using the slow rotation approximation.
The expression of the timescales is adopted from Lindblom et al.~(1999).
It is noted that the shear viscosity term (Eq.(10)) is evaluated in our
analysis by numerical differentiation of the corresponding eigenfunction.
This leads to the relative error of $\sim 10$ \% in the shear viscous 
timescale $\tau_s$ for the present mesh size ($32$ and $11$ in the r
and $\theta$ direction).

Together with it plotted are the curves on which instability time-scale 
is equal to the cooling time-scale of the star. If we assume the `standard' 
modified URCA process \cite{shapiro-teukolsky} to be dominant in the initial 
stage of neutron star cooling,\footnote{Though nucleon 
bremsstrahlung is also one of the standard cooling channels, its contribution 
is order of magnitude smaller than that of the modified URCA process. See Friman \& Maxwell~\shortcite{friman-maxwell}.} the effective cooling time-scale 
$\tau_{\rm cool}$ at the temperature $T$(K) is defined by:
\begin{equation}
\tau_{\rm cool} = \left[\frac{d\ln T_9}{dt}\right]^{-1} = \frac{6 t_c}{T_9^6},
\end{equation}
where $T_9=T/10^9$ and $t_c$ is a constant characterizing the cooling time 
which is typically $\sim 1$y for this process.

Apart from the early phase, hot young neutron stars may evolve along this 
evolutionary curve $\tau_r=\tau_{\rm cool}$. Suppose that a neutron star is 
born at the temperature $\sim 10^{11}$K and with the rotational frequency 
above the bottom of the critical curve. At the beginning it is stable against 
the r-mode instability and evolves along the horizontal
line on the $T-f$ plane. In a few second, it enters the unstable region
against the r-mode and the initial (stochastic) perturbation begins to grow.
As long as $\tau_r>\tau_{\rm cool}$, it evolves almost horizontally. Once 
it reaches the `evolution curve', the star begins to go down along this curve,
since any hypothetical displacement of the star from this curve will be 
amended by the growth of instability or the cooling.  If the star is located 
at the right side of this curve, the star will evolve nearly horizontally 
leftwards due to rapid cooling. If the star is located at the left side of 
this curve, the star will evolve nearly vertically downward due to rapid
loss of gravitational waves. 

When the star goes down along the evolution curve and approaches the 
critical curve asymptotically, the instability ceases to work effectively 
and the star will not be spun down by the instability any more. The 
location of this evolution curve in the $T-f$ plane is insensitive to the value of constant $t_c$.
Consequently every neutron star seems to settle down to an almost universal 
terminal rotational frequency, i.e. $\sim 100$Hz at the end of the spin 
evolution. If $t_c=1$y, this frequency is 
$90$Hz and is reached when the temperature of the star equals to 
$1.1\times 10^9$K.

On the other hand, if some exotic rapid cooling mechanisms dominate, 
the picture may be changed drastically. In Fig~\ref{fig2}, an evolution curve 
defined by $\tau_r=\tau_{\rm cool}$ for the model with pion condensation
is also plotted (dashed curve). The pion condensation enhances the cooling 
because a new process analogous to the ordinary URCA process 
\cite{shapiro-teukolsky} begins to operate.  The effective cooling time-scale 
for this process is written as follows:
\begin{equation}
  \tau_{\rm cool} = \frac{4 t'_c}{T_9^4},
\end{equation}
where $t'_c\sim 200$s. The evolution curve of $\tau_r=\tau_{\rm cool}$ shifts 
upward and the terminal rotational frequency also shifts to about $200$Hz.

%% Apr.23
Commenting on an evolution of a star by a rapid cooling mechanism may
be appropriate here.
By an exotic cooling mechanism such as the pionic reaction, 
the core fluid of a hot and young neutron star cools rapidly 
down to $\sim 10^8$ K, while the crust component of it remains 
hot ($\sim 10^9$K) in the first few decades (see Lattimer et al.~(1994) 
for instance). 
Then the standard assumption in the r-mode instability that the stellar 
temperature remains uniform is broken.
Andersson et al.~\shortcite{andersson-kokkotas-schutz} argues that the warm
crust work as a heat reservoir and the rapid cooling effect on the
stellar evolution by the r-mode is moderated. This reduces the difference
between the outcome of the rapid and the standard cooling.
We further note that the viscous heating by the r-mode shear motion 
\cite{levin}, which is omitted here and in the other works dealing 
with hot and young neutron stars, may delay the cooling and have the 
same effect as the crust. This heating prolongs the lapse during
which the r-mode instability is effective ~(Yoshida, unpublished).

This means that in order to know the significant effect of the r-mode 
instability on spin evolution of newly born hot neutron stars, we need
to know more about the nature of dense matter at high and intermediate 
temperatures.

%% Apr.23
\section{Summary}
We show the first results of the numerical mode analysis of the classical
r-mode for the {\it rapidly rotating and compressible} stellar models.
The eigenfrequencies and their eigenfunction of the rapidly rotating
stellar models are numerically computed nearly to the Kepler limit.
The time scale of the r-mode instability is evaluated by the eigenfrequency
and the eigenfunction.

The behaviour of the eigenfrequency is confirmed to be well approximated
by the third order expansion in the angular frequency $\Omega$.
The difference between the critical curves drawn by the full computation 
and by the slow rotation approximation is shown to be rather small.
This certificates the use of the slow rotation approximation
in the third order to investigate the r-mode instability.
%%

%-------------------- bibliographies --------------------------------

\end{document}